\documentclass[12pt]{article}
\usepackage{latexsym}
\usepackage{amsmath}
\usepackage{amsfonts}
\usepackage{amssymb}
 
\newcommand{\mR}{{\mathbb R}}
\newcommand{\mH}{{\mathcal H}}
\newcommand{\mA}{{\mathcal A}}
\title{Nonlinear realizations,  the orbit method  and Kohn's theorem}
\author{K. Andrzejewski,
J. Gonera, P. Kosi\'nski\thanks{e-mail: pkosinsk@uni.lodz.pl}\\
\small Department of Theoretical Physics and Computer Science, \\
\small University of \L\'od\'z,\\
\small Pomorska 149/153, 90-236 {\L}\'od\'z, Poland
}
\date{}
\begin{document}
\maketitle
\begin{abstract}
The orbit method is used to describe the centre of mass motion of the system of particles with fixed charge to mass ratio moving in homogeneous magnetic field and confined by harmonic potential. The nonlinear action of symmetry group on phase space is identified and compared with the one obtained with the help of Eisenhart-Duval lift.
\end{abstract}
\section{Introduction}
In the recent  paper \cite{b1} Gibbons and Pope considered the system of nonrelativistic particles with fixed charge to mass ratio, interacting through  the potential depending on their mutual separation, which move in the uniform magnetic field and confined by harmonic external potential. They gave a group--theoretical interpretation of Kohn's theorem \cite{b2} concerning the possibility of separating the centre of mass motion. Using old Larmor result one can eliminate the magnetic field  at the expense of modifying the external harmonic potential. It appears then that the Kohn's theorem can be ascribed to the invariance of the system under the action of Newton-Hooke  group in 2+1 dimensional nonrelativistic spacetime (Zhang and Horvathy \cite{b3} pointed out, however,  that this symmetry is an "imported" Galilean symmetry  resulting from applying the Niederer's transformation \cite{b4}).
\par
Gibbons and Pope made an interesting observation concerning the symmetry of the centre of mass dynamics. By performing the Eisenhart-Duval lift \cite{b5} of the system they found that the resulting metrics provides pp-wave solution of Einstein-Maxwell equations \cite{b6} which, in fact, represents bi-invariant metric on the universal covering of centrally extended  $E(2)$ group (the so-called Cangemi-Jackiw group CJ \cite{b7}). The isometry group of this metric is $(CJ\times CJ)/N$, $N$ being the diagonal element of the centre. 
\par In the present note we study this problem in some more detail. Using the method of orbits \cite{b8} we give the group--theoretical interpretation of the centre of mass phase space and compare our construction with that of Gibbons and Pope. We also show that the dynamics exhibiting Newton-Hooke symmetry can be described in terms of nonlinear realization \cite{b9}  in a similar way as  proposed by Ivanov, Krivonos and Leviant \cite{b10} in the case of conformally invariant dynamics.
\section{Dynamics and symmetry of planar oscillator}
Our starting point is the Lie algebra of the $(CJ\times CJ)/N$ group \cite{b1}. 
\begin{equation}
\label{e1}
\begin{array}{l}
[L,L_i]=i\epsilon _{ij}L_j, \quad [L_i,L_j]=i\epsilon _{ij}T, \\  
\left[R,R_i\right]=-i\epsilon _{ij}R_j, \quad [R_i,R_j]=-i\epsilon _{ij}T. 
\end{array}
\end{equation}
It can be integrated to topologically trivial ($\sim \mR^7$) group  with composition law
\begin{align}
\label{e2}
&(a_{Li},u_L,a_{Ri},u_R,v)*(a_{Li}',u_L',a_{Ri}',u_R',v')=
\Big(a_{Li}+a_{Li}'\cos u_L+\epsilon_{ij}a_{Lj}'\sin u_L, \nonumber\\
&u_L+u_L',a_{Ri}+a_{Ri}'\cos u_R-\epsilon_{ij}a_{Rj}'\sin u_R,
u_R+u_R',v+v'+\\
& \frac{1}{2}\epsilon_{ij}(-a_{Li}a_{Lj}'\cos u_L+a_{Ri}a_{Rj}' \cos u_R)+\frac{1}{2}(a_{Li}a_{Li}'\sin u_L+a_{Ri}a'_{Ri}\sin u_R)\Big).\nonumber
\end{align}
The algebra (\ref{e1}) in nothing but the Newton-Hooke algebra. In fact, defining new generators
\begin{align}
\label{e3}
P_i&=\frac{1}{\sqrt 2}(L_i+\epsilon_{ij}R_j),\nonumber\\
B_i&=\frac{1}{\sqrt2}(\epsilon_{ij}L_j+R_i),\nonumber\\
H&=L+R,\\
J&=L-R,\nonumber\\
M&=-T,\nonumber
\end{align}
we verify that they obey the commutation rules of Newton-Hooke algebra.
\par Assuming that our symmetry acts transitively (which is the case for centre of mass variables) we can use the orbit method \cite{b8}. To this end consider the linear space dual to the Lie algebra (\ref{e1}) parameterized as follows
\begin{equation}
\label{e4}
X=\vec {\xi}_L\tilde{\vec{ L}}+{\zeta }_L\tilde{{ L}}+\vec {\xi}_R\tilde{\vec{ R}}+{\zeta}_R\tilde{{R}}+\mu\tilde T,
\end{equation}
where $\tilde{ \vec {L}},\tilde L$ etc. denote the elements of dual basis. For our purposes it is sufficient to consider the coadjoint action of one-parameter subgroups. It reads:
\par
$g(\vec{a}_L)=e^{i\vec{a}_L\vec L}:$
\begin{align}
\label{e5}
\xi_{Li}'&=\xi_{Li}-\epsilon_{ik}a_{Lk}\mu,\nonumber\\
\zeta_L'&=\zeta_L+\epsilon_{ik}a_{Lk}\xi_{Li}-\frac{1}{2}\vec{a}_L^2\mu,\nonumber\\
\xi_{Ri}'&=\xi_{Ri}, \\
\zeta_R'&=\zeta_R,\nonumber\\
\mu'&=\mu,\nonumber
\end{align}
\par
$g(u_L)=e^{iu_L L}:$
\begin{align}
\label{e6}
\xi_{Li}'&=\xi_{Li}\cos u_L+\epsilon_{ik}\xi_{Lk}\sin u_l,\nonumber\\
\zeta_L'&=\zeta_L,\nonumber\\
\xi_{Ri}'&=\xi_{Ri} ,\\
\zeta_R'&=\zeta_R,\nonumber\\
\mu'&=\mu,\nonumber
\end{align}
\par
$g({u}_R)=e^{iu_R R}:$
\begin{align}
\label{e7}
\xi_{Li}'&=\xi_{Li},\nonumber\\ 
\zeta_L'&=\zeta_L,\nonumber\\
\xi_{Ri}'&=\xi_{Ri}\cos u_R-\epsilon_{ik}\xi_{Rk}\sin u_r, \\
\zeta_R'&=\zeta_R,\nonumber\\
\mu'&=\mu,\nonumber
\end{align}
\par
$g(\vec{a}_R)=e^{i\vec{a}_R\vec R}:$
\begin{align}
\label{e8}
\xi_{Li}'&=\xi_{Li},\nonumber\\
\zeta_L'&=\zeta_L,\nonumber\\
\xi_{Ri}'&=\xi_{Ri}+\epsilon_{ik}a_{Rk}\mu, \\
\zeta_R'&=\zeta_R\nonumber-\epsilon_{ik}a_{Rk}\xi_{Ri}-\frac{1}{2}\vec{a}_R^2\mu,\\
\mu'&=\mu.\nonumber
\end{align}
Let us note that $\mu$ is an invariant of coadjoint action; we shall assume that $\mu\neq 0$ and, in view of eqs. (\ref{e3}), $\mu=-m$, $m>0$ (the case $\mu>0$ can be dealt with analogously). It follows form eqs. (\ref{e5})-(\ref{e8}) that any orbit with $\mu \neq 0$ contains the point corresponding to $\vec{\xi}_L=0$, $\vec{\xi}_R=0$; denote by $\lambda$ and $\rho$ the coordinates $\zeta_L$, resp. $\zeta_R$ for this point. Using again eqs. (\ref{e5})-(\ref{e8}) we find the parameterization of arbitrary  orbit (with $\mu\neq 0$):
\begin{equation}
\label{e9}
\left(\vec {\xi}_L,\vec{\xi}_R,\zeta_L=\lambda+\frac{{\vec {\xi}_L}^2}{2m},\zeta_R=\rho+\frac{\vec {\xi}_R^2}{2m},m\right).
\end{equation}
The orbit is characterized by the choice of $m,\lambda$ and $\rho$; note that the invariants $\lambda$ and $\rho$ correspond to the Casimir operators
\begin{equation}
\label{e10}
C_L=TL-\frac{1}{2}\vec{L}^2,\quad C_R=TR-\frac{1}{2}\vec{R}^2.
\end{equation}
According to the general theory \cite{b8} one can write out the relevant Poisson brackets. Before doing that we make a redefinition of basic variables suggested by eqs. (\ref{e3}):
\begin{align}
\label{e11}
q_i=\frac{1}{m\sqrt2}(\epsilon_{ij}\xi_{Lj}+\xi_{Ri}), \quad
p_i=\frac{1}{\sqrt2}(\epsilon_{ij}\xi_{Rj}+\xi_{Li}).
\end{align}
Then the only nontrivial Poisson bracket reads
\begin{equation}
\label{e12}
\{q_i,p_k\}=\delta_{ik},
\end{equation}
whereas, denoting by $h$ (resp. $j$) the function on phase space corresponding to the generator $H$ (resp. $J$),
\begin{equation}
\label{e13}
\begin{array}{c}
h=\lambda+\rho+\frac{\vec{p}^2}{2m}+\frac{m\vec{ q}^2}{2},\\ 
j=\lambda-\rho+q_1p_2-q_2p_1.
\end{array}
\end{equation}
We see that $\lambda+\rho$, ($\lambda-\rho$) play a role of internal energy (angular momentum). 
\par The structure of phase space can be also  described from the point of view of nonlinear realizations \cite{b9}. Call $\mH=\{L,R,T\}$, $\mA=\{\vec{L},\vec{R}\}$. Then $\mH$ is the stability algebra of our distinguished point on the orbit. Moreover, $[\mH,\mH]\subset \mH$, $[\mH,\mA]\subset \mA$, and $[\mA,\mA]\subset\mH$ so the symmetry group acts on phase space according to the nonlinear realization linearizing  on the subgroup generated by $L,R$ and $T$; $\vec{\xi}_L$  and $\vec{\xi}_R$ are the preferred variables, in the standard terminology of nonlinear realizations. $\vec{\xi}'s$ are simply related to the variables in exponential parameterization. Let the elements of the coset space be parameterized as $w=\exp i(\vec {z}_L\vec{L}+\vec{z}_R\vec{R})$. Then, as one easily checks
\begin{equation}
\label{e14}
z_{L,Ri}=\frac{1}{\mu}\epsilon_{ij}\xi_{L,Rj}.
\end{equation}
Let us compare the above action  of $(CJ\times CJ)/N$ on the phase space  with that defined by Gibbons and Pope \cite{b1}. There $CJ\times CJ$ acts on $CJ$ by left and right multiplication. The Lie algebra of $CJ\times CJ$ is given by eqs. (\ref{e1}) except that one has to distinguish left and right $T's$:
\begin{equation}
\label{e15}
[L_i,L_j]=i\epsilon_{ij}T_L,\quad [R_i,R_j]=-i\epsilon_{ij}T_R.
\end{equation}
Now, the action of $CJ\times CJ$ on $CJ$  is the nonlinear realization  of $CJ\times CJ$ corresponding to the following decomposition of its Lie algebra:
\begin{equation}
\label{e16}
\begin{array}{c}
\tilde\mH=\{L-R,\vec{L}-\vec{R},T_L-T_R\},\\
\tilde\mA=\{L+R,\vec{L}+\vec{R},T_L+T_R\},
\end{array}
\end{equation}
(in particle physics terminology we are considering chiral symmetry broken down to  the vector one \cite{b9}; cf. also Refs. \cite{b9A}). Our group is, however, $(CJ\times CJ)/N$; this is taken into account by identifying $T_L=T_R$. Therefore, we conclude that the nonlinear realization of $(CJ\times CJ)/N$ considered in Ref. \cite{b1} corresponds to the choice of $E(2)$ subgroup generated by $L-R$ and $\vec{L}-\vec{R}$ as the one on which the action linearizes. Note that the coset space $((CJ\times CJ)/N)/E(2)$ spanned by $\tilde A$ is the group manifold of $CJ$ but the group structure is not generated by the  elements of $\tilde A$; this is again typical for chiral symmetries.
\par
Finally we note that our dynamics admits a description similar to that use by Ivanov, Krivonos and Leviant in the case of conformal mechanics \cite{b11}. Using the orbit method it is not difficult to generalize their method to other dynamical systems \cite{b11,b12}. The relevant equations reads
\begin{equation}
\label{e17}
e^{itH}w(t)=w(0)h(t),
\end{equation}
where $h(t)$ are elements of stability subgroup of distinguished point on coadjoint orbit while $w(t)$ belong to the relevant coset manifold. In our case, using eqs. (\ref{e11}) and (\ref{e14}) together with $H=L+R$ we find
\begin{equation}
\label{e18}
q_i(t)=q_i(0)\cos t+\frac{p_i(0)}{m}\sin t,
\end{equation}
in agreement with eqs. (\ref{e12}) and (\ref{e13}).
\section{Conclusions}
Using the orbit method we found the description of the phase space of center of mass variables in terms of geometry of Cangemi-Jackiw group. It provides an alternative  to the approach described by Gibbons and Pope and based on the notion Eisenhart lift.
\par
In both schemes the main role is played by the action of  $(CJ\times CJ)/N$ group on the trivial manifold ($\mR^4$) of Cangemi-Jackiw group. The corresponding nonlinear realizations are, however, different. This is not surprising taking into account different  interpretations of group manifold. In our case it is directly identified with phase space. This one of the main advantages of the orbit method. 
Let us finally note that the orbit method has been recently considered in the context of dynamics on noncommutative phase space \cite{b13}.
\par
{\bf Acknowledgments}
We are grateful to Prof. P. Horvathy for remarks.
This work  is supported  in part by  MNiSzW grant No. N202331139

\end{document}